\documentclass[%
preprint,
 amsmath,amssymb,
 aps,
]{revtex4-2}
\newcommand{\klow}{{k_{\rm low}}}
\newcommand{\kh}{{k_{\rm high}}}

\newcommand{\red}{\color{red}}
\def\red#1 {\textcolor{red} {#1} }

\usepackage{physics}
\usepackage{array}
\usepackage{setspace}
\newcolumntype{C}[1]{>{\centering\arraybackslash}m{#1}}
\usepackage{xcolor}
\usepackage{graphicx}% Include figure files
\usepackage{dcolumn}% Align table columns on decimal point
\usepackage{bm}% bold math
\usepackage{subcaption}
\begin{document}

\preprint{APS/123-QED}

\title{Sparse-Stochastic Fragmented Exchange for Large-Scale Hybrid TDDFT Calculations}
% Force line breaks with \\

% \author{Ann Author}
%  \altaffiliation[Also at ]{Physics Department, XYZ University.}%Lines break automatically or can be forced with \\

% \collaboration{MUSO Collaboration}%\noaffiliation

\author{Mykola Sereda*}
%\email{koyla@ucla.edu}
\affiliation{Department of Chemistry and Biochemistry, University of California, \\ Los Angeles, California, 90095, USA}

\author{Tucker Allen*}
%\email{tuckerallen27@ucla.edu}
\affiliation{Department of Chemistry and Biochemistry, University of California, \\ Los Angeles, California, 90095, USA}

\author{Nadine C Bradbury}
\affiliation{Department of Chemistry and Biochemistry, University of California, \\ Los Angeles, California, 90095, USA}

\author{Khaled Z Ibrahim}
 % \email{kzibrahim@lbl.gov}
\affiliation{Computer Science Department, Lawrence Berkeley National Laboratory,
One Cyclotron road, Berkeley, CA 94720, USA}

\author{Daniel Neuhauser} 
 \email{dxn@ucla.edu} 
 \affiliation{Department of Chemistry and Biochemistry, and California Nanoscience Institute, UCLA, Los Angeles CA 90095-1569 USA}

% \collaboration{CLEO Collaboration}%\noaffiliation

\date{\today}% It is always \today, today,

\begin{abstract}
We extend our recently developed sparse-stochastic fragmented exchange formalism for ground-state hybrid DFT (ngH-DFT) to calculate absorption spectra within linear-response time-dependent Generalized Kohn-Sham DFT (LR-GKS-TDDFT), for systems consisting of thousands of valence electrons within a grid-based/plane-wave representation. A mixed deterministic/fragmented-stochastic compression of the exchange kernel, here using long-range explicit exchange functionals, provides an efficient method for accurate optical spectra. Both real-time propagation as well frequency-resolved Casida-equation-type approaches for spectra are presented, and the method is applied to large molecular dyes.  
\end{abstract}

\maketitle

%\tableofcontents

\section{Introduction}
Time-dependent density-functional theory (TDDFT) \cite{casida_1995} has been extensively used for predicting photoelectron spectra \cite{Casida_1996, yabana_time-dependent_1999, yabana_optical_1997,lopata_linear-response_2012, debeer_george_time-dependent_2008}, materials screening, optoelectronic device design \cite{lerme_optical_1998, qian_time-dependent_2006, baer_ab_2003, baer_method_2001, zelovich_state_2014}, and nonlinear optical response.~\cite{ding_efficient_2013, parker_quadratic_2018}
A common approximation for the exchange-correlation kernel is the local density approximation (LDA) which in DFT is notorious for underestimating the band gap in periodic compounds and the ionization potential in finite systems.~\cite{perdew_density_2009} Similarly, in time-dependent calculations, the commonly used adiabatic LDA (ALDA) approximation often gives incorrect absorption peaks and shifted cutoff regions in higher-harmonic generation spectra. More sophisticated methods, such as the Bethe-Salpeter Equation (BSE), TD-CIS(D), or TD-CC2, give a better description of electron correlation and result in a more accurate description of electron dynamics in both the linear-response and strong field regime.~\cite{li_real-time_2020} However, these methods can scale poorly with system size, and significant computational or theoretical developments are still needed to address large systems. 

Hybrid functionals in DFT admix exact exchange with an approximate exchange-correlation (XC) functional.~\cite{Becke1993exact} Inclusion of exact exchange helps improve bandgaps and alleviate the self-repulsion in the KS potential.~\cite{baer_tuned_2010} Global hybrids, such as PBE0 and B3LYP, include a fractional amount of exchange at all length scales.~\cite{HERBERT2023} On the other hand, range-separated hybrid (RSH) functionals partition the Coulomb interaction into a short-range component, treated locally, and a long-range part treated explicitly. The strategy is to correct the asymptotic behavior of the KS potential while maintaining a local (or semi-local) XC potential description at short-range.~\cite{Tawada_2004, baer_density_2005} 

Popular RSH's, such as CAM-B3LYP, use a fixed range-separation parameter, here denoted by $\gamma$, to partition the short- and long-range components of the exchange.~\cite{savin1997, YANAI200451} Optimally-tuned (OT)-RSH's obtain $\gamma$ by enforcing a physical condition on the system of interest based on first-principles.~\cite{baer_density_2005} OT-RSH functionals recover the correct $-1/r$ asymptotic behavior of the exchange potential, essential for an accurate description of charge-transfer excitations and electron-hole bound states.~\cite{baer_tuned_2010,vlvcek2019stochastic} In this work, we use the Baer-Neuhauser-Lifshitz (BNL) OT-RSH functional.~\cite{baer_density_2005} For this functional, the range-separation parameter $\gamma$ is obtained by enforcing piece-wise linearity of the total energy with electron number.~\cite{baer_tuned_2010}

For excited states, ALDA-TDDFT simplifies the time-dependent many-body problem to an independent orbital propagation. Use of a hybrid functional retains this independent orbital propagation and significantly improves the ailments of ALDA. Previous TDGKS studies with the BNL functional accurately described the vertical excitation energies of a range of acenes, achieving excellent agreement with experimental and coupled cluster values.~\cite{lopata2011} Further, an optimally-tuned BNL functional was able to  correctly predict charge-transfer excitations in light-harvesting molecules.~\cite{bnl_tddft_baer2011,baer_bnltddft_2012} However, this incurs a steep computational scaling. 

Methods to reduce the computational scaling within a given TDGKS implementation depend on the choice of basis-set. (We use below the terms TDGKS and hybrid-TDDFT interchangeably.) In atomic-orbital (AO) basis-sets, efforts to reduce the cost of exact exchange include density fitting, Cholesky decomposition, resolution of the identity approaches \cite{aquilante_low-cost_2007,headgordon_2015, qin2020_densityfitting,rhofitting_frozoni,sharma2022}, and spatial localization methods.~\cite{Liu2015, distasio2021}. In plane-wave (PW) representations, projection methods have been developed that circumvent the use of virtual orbitals.~\cite{Hutter_2003,natan_projectors2016} Adaptively compressed exchange, a low-rank decomposition method, has significantly reduced the cost of exchange and has been applied to excited states.~\cite{Liu2022} Diagonal hybrid TDDFT has also been implemented which balances explicit diagonal exchange elements in the Casida matrix with off-diagonal elements approximated by an LDA kernel.~\cite{hybrid_diag2020,hybrid_diag2_2023}

In our previous work, we developed a real-time, grid-based TDGKS algorithm that reduces the scaling of exact exchange by a stochastic decomposition of the time-dependent density matrix and the Coulomb kernel.~\cite{vlvcek2019stochastic} A similar stochastic approach in the frequency domain has also been developed.~\cite{stochastic_rsh_Neuhauser_2020} 

Here we present an alternate approach with very little stochastic error by introducing a mixed deterministic/sparse-stochastic sampling of the Coulomb kernel in $k$-space and avoiding a stochastic resolution of the density matrix. The numerically-large low-$k$ components of the exchange kernel are treated deterministically and the many numerically-small high-$k$ elements are represented through a sparse-stochastic basis. The size of this sparse high-$k$ basis becomes constant, irrespective of system size.~\cite{neargap_2023} In addition, we work in a valence-conduction subspace of Kohn-Sham (KS) orbitals, nearest to the Fermi level.

When combining these techniques we extract readily, with modest costs, the time-dependent properties of large systems. In the following sections, we outline how the sparse-fragmented exchange method is applied to linear-response calculations in both the real-time and frequency domains. The method is applied to naphthalene, fullerene, a chlorophyll a (Chla) dye monomer, and a hexamer chlorophyll dye complex found at the reaction center of Photosystem II (RC-PSII). ~\cite{Chlagasphase_original_2019, Forster_ADF_Chromo_2022} The latter system is large (1320 valence electrons) and therefore very expensive for traditional deterministic approaches. 

\section{Theory}
For a small external perturbation of an electric field in a given direction, taken as $\hat{x}$ here for simplicity, we evaluate the dipole-dipole correlation function through either a real-time propagation (Sec. A) or a frequency-domain linear-response approach (Sec. B). The two approaches have each their specific benefits. The real-time approach is simple to derive and easily extendable to strong and continuous electric field pulses. On the other hand, the frequency-domain method is numerically faster and specifically optimized for extracting the linear-response absorption spectra.

Since we go beyond the Tamm-Dancoff approximation (TDA) for the frequency-domain approach, the two methods yield essentially identical results for the linear-response optical spectra. For completeness, we overview the mixed deterministic/stochastic fragmentation of the Coulomb kernel in Sec. C, and for further details refer readers to Ref.~\cite{neargap_2023}.

\subsection{Hybrid TDDFT in the Valence-Conduction Subspace}

The aim of the real-time approach is to efficiently solve the TDSE (time-dependent Schrodinger equation),
\begin{equation}
    i \partial_t \ket{\psi_i} = \hat{h}[n(t),\rho(t)]\ket{\psi_i},
\end{equation} 
in the TDGKS framework and predict the optical gap in the absorption spectrum.

The starting point comes from a near-gap Hybrid DFT (ngH-DFT) ground-state calculation, where the GKS ground-state molecular orbitals (MOs) are represented as a rotation matrix of the LDA ground state MOs \cite{neargap_2023}
\begin{equation}
    \ket{\psi_i(t)} = \sum_{p}C_{pi}(t) \ket{\phi_p},
\end{equation}
where the basis set is the subset of eigenfunctions of the ground-state Kohn-Sham DFT LDA Hamiltonian:
\begin{equation}
    \hat{H}\phi_s = \varepsilon_s \phi_s.
\end{equation}
The molecular orbitals, $\phi_p$, are divided into four sets of states: $N_{core}$ core, $N_v = N_{occ}-N_{core}$ valence (occupied), $N_{c}$ conduction (virtual), and the remaining high-lying conduction states. We work in the combined valence and conduction subspace, $N_v \oplus N_c$. The effect of the high-lying conduction subspace is not considered and the effect of the core states is approximated as a perturbative scissors correction, discussed in Eq. (\ref{eq:xcore}). Throughout this paper, the indices $p,q,...$ run over the combined $M = N_v + N_c$ near-gap states, $i,j,...$ refer to valence orbitals, and $a,b,...$ to conduction orbitals. 

In this basis, the GKS Hamiltonian is represented as
\begin{equation}
    H_{pq}(t) = \bra{\phi_p}\hat{h}(t)\ket{\phi_q},
\end{equation}
which is evaluated as
\begin{equation}
    H_{pq}(t) = \bra{\phi_p}\hat{h}_0 + \delta \hat{v}^{\gamma}[n(t)] + \hat{X}^{\gamma}_{val}[\rho(t)] + \hat{X}^{\gamma}_{core}\ket{\phi_q}.
\end{equation}
As mentioned, the range-separation parameter $\gamma$ is non-empirically obtained by enforcing the ionization potential theorem; further details are provided in Refs.~\cite{baer_tuned_2010,neargap_2023} The time-independent part of the Hamiltonian, $\hat{h}_0$, is
\begin{equation}
    \hat{h}_0 = -\frac{1}{2}\hat{\nabla}^2 + \hat{v}^{NL}_{eN} + \hat{v}^{\gamma}[n(t=0)],
\end{equation}
which includes the Kohn-Sham potential
\begin{equation}
    \hat{v}^{\gamma}[n(t)] = \hat{v}_{eN}^{local} + \int\frac{n(r',t)}{|r-r'|}dr' + \hat{v}_{XC}^{SR,\gamma}[n(t)].
\end{equation}
Here $\hat{v}_{XC}^{SR,\gamma}[n(t)]$ is the local short-range exchange correlation potential.
The next term in the Hamiltonian is the difference between the short-range Kohn-Sham potential at time $t$ and at $t=0$,
\begin{equation}
    \delta \hat{v}^{\gamma} [n(t)] = \hat{v}^{\gamma}[n(t)] - \hat{v}^{\gamma}[n(t=0)].
\end{equation}
The core contribution to the density is always approximated as time-independent,
\begin{equation}
    n(r,t) = n_{core}(r) + n_{val}(r,t).
\end{equation}
The long-range exchange matrix elements are calculated at each time-step,
\begin{equation}\label{eq:Xelements}
\bra{\phi_p}\hat{X}^{\gamma}_{val} \ket{\phi_q} = -\sum_{st} P_{st}(t) (\phi_q \phi_s |u^{\gamma}|\phi_t \phi_p),
\end{equation}
where the long-range Coulomb kernel is $u^\gamma(|r-r'|)=\text{erf}(\gamma|r-r'|)/|r-r'|$, so assuming real orbitals,
\begin{equation}
\label{eq:ugamma_mat}
(\phi_q\phi_s|u^\gamma|\phi_t\phi_p)= \int \phi_q(r)\phi_s(r)u^\gamma(|r-r'|)\phi_t(r')\phi_p(r')dr'dr.
\end{equation}
To circumvent the prohibitive cost of calculating the exchange matrix elements in Eq. (\ref{eq:ugamma_mat}), the long-range Coulomb kernel is split into two parts, one treated deterministically and one that we accurately represent through a sparse-stochastic basis. Using a stochastic resolution of the identity for the Coulomb kernel in the reciprocal space (see Eq. (\ref{identity})), the exact exchange matrix elements are
\begin{equation}
    \label{eq:xval_sto}
    \bra{\phi_p}\hat{X}^{\gamma}_{val} \ket{\phi_q} = -\sum_{st} P_{st}(t) \bra{\phi_q \phi_s}\ket{\xi}\bra{\xi}\ket{\phi_t \phi_p},
\end{equation}
where $\{|\xi\rangle\}$ is a small set of auxiliary vectors, as detailed in Sec. C.  
Eq. (\ref{eq:xval_sto}) is efficiently computed as
\begin{equation}
\bra{\phi_p}\hat{X}^{\gamma}_{val} \ket{\phi_q} = -\sum_{\xi i} u_{p\xi i}(t) u^*_{q\xi i}(t),
\end{equation}
where we only have to store the vectors (see Sec. C)
\begin{equation}
    u_{p\xi i}(t) = \sum_t C_{ti}(t)\bra{\xi}\ket{\phi_t \phi_p}.
\end{equation}

Next, the effect of the core states on the exact exchange operator, $\hat{X}^{\gamma}_{core}$, is approximated as a time-independent scissors correction, 
\begin{equation}
    \label{eq:xcore}
    \hat{X}^{\gamma}_{\text{core}} =  \hat{P}d\epsilon_{\gamma,\text{HOMO}} + \hat{Q}d\epsilon_{\gamma,\text{LUMO}},
\end{equation}
where $\hat{P}$ projects into the $N_v$ subspace, $\hat{Q}$ projects into the $N_c$ subspace, and
\begin{equation}
  d\epsilon_{\gamma,a} = 
 \sum_{f \in core} (\psi_{a} \phi_f|u^{\gamma}|\phi_f \psi_{a}).
\end{equation}

By inserting the GKS wavefunction, expanded in the LDA MO basis, into the TDSE, and using the orthogonality of the LDA MOs, an ordinary basis set TDSE is obtained
\begin{equation}
    i\partial_t C(t) = H(t)C(t).
\end{equation}
We excite the initial state with a small perturbation, of strength $\Delta$,
\begin{equation}
\label{strength}
    C(t=0^+) = e^{-i\Delta D}C(t=0),
\end{equation}
where
\begin{equation}
    D_{pq} = \bra{\phi_p} \hat{x} \ket{\phi_q}.
\end{equation}
The time-dependent coefficients are propagated using the midpoint rule for the time evolution operator
\begin{equation}
    C(t+dt) \simeq e^{-idtH(t+\frac{dt}{2})}C(t),
\end{equation}  
where we use a linear extrapolation approach,
$H(t+\tfrac{dt}{2}) = 1.5H(t) - 0.5H(t-dt)$.

The dipole moment along $\hat{x}$ is extracted,
\begin{equation}
    \mu (t) = \sum_{qp} P_{qp}(t) D_{pq}.
\end{equation}
The absorption spectra is then generated from the time-dependent dipole moment as usual \cite{li_real-time_2020}
\begin{equation}
    S(\omega) = \frac{1}{3} \text{Tr}[\sigma(\omega)],
\end{equation}
where
\begin{equation}
\label{eq:sig}
    \sigma_{ii}(\omega) = \frac{4 \pi \omega}{c} \text{Im}[\alpha_{ii}(\omega)],
\end{equation}
and the polarizability is calculated from $\alpha_{ii}(\omega) = \frac{\mu_i(\omega)}{\Delta}$.

\subsection{$\sigma(\omega)$ via an iterative Chebyshev procedure}

The frequencies of the TDSE solve an eigenvalue equation, known as the Casida equation.~\cite{casida_1995} For reference, see a traditional derivation including Fock exchange in Ref.~\cite{Negele1982}. The Casida equation is formally
\begin{equation}
    \mathcal{L}\begin{pmatrix} f^+\\ f^- \end{pmatrix} = \hbar\omega \begin{pmatrix}1&0\\0&-1\end{pmatrix}\begin{pmatrix} f^+\\ f^- \end{pmatrix},
\end{equation}
where the Liouvillian is
\begin{equation}
    \mathcal{L}=\begin{pmatrix}A&B\\-B&-A\end{pmatrix},
\end{equation}
and $f^+$ and $f^-$ refer to the orbitals of the linear-response perturbation. In the space of single-particle singlet excitations, the orbital Hessians \emph{A} and \emph{B} are
\begin{equation}
\begin{split}
    &A(ia;jb)=(\varepsilon_a-\varepsilon_i)\delta_{ij}\delta_{ab}+2(ia|jb)+(ia|f_{\text{XC}}|jb)-(\phi_a\phi_b|u^\gamma|\phi_i\phi_j)\\
    \\
    &B(ia;bj) = 2(ia|bj)+(ia|f_{\text{XC}}|bj) - (\phi_a\phi_j| u^\gamma|\phi_i \phi_b),
\end{split}
\end{equation}
with two-electron integrals 
\begin{equation}
    (ia|jb)=\int \phi_i(r)\phi_a(r)|r-r'|^{-1}\phi_j(r')\phi_b(r')dr'dr,
\end{equation} so $(ia|jb)$=$(ia|bj)$. Similarly to the real-time approach presented in Sec. A, the XC kernel elements are computed within an attentuated ALDA scheme to first order.~\cite{RT-TDDFT_Roi2004} 

To track the response of the system to an electric field in the \emph{x}-direction one uses an initial excitation spinor composed of two dipole matrix elements
\begin{equation}
    |\chi\rangle = \begin{pmatrix} \chi_{ia}^+\\ \chi_{ia}^- \end{pmatrix}
    =\begin{pmatrix} +\langle{\phi_a}| x | {\phi_i}\rangle\\\
    -\langle{\phi_a}| x | {\phi_i}\rangle\end{pmatrix}.
\end{equation}
Application of the Liouvillian matrix on a general vector $(f^+,f^-)^T$ is then 
\begin{equation}
\label{Lf_pos2}
    (\mathcal{L}f^+)_{ia}=(\varepsilon_a-\varepsilon_i)f^+_{ia}+\langle\phi_a|\delta v_H|\phi_i\rangle+\langle\phi_a|\delta v^\gamma_{XC}|\phi_i\rangle -\langle \phi_a|y^+_i\rangle-\langle \phi_a|z^-_i\rangle,
\end{equation}
and for backward transitions,
\begin{equation}
\label{Lf_neg2}
    (\mathcal{L}f^-)_{ia}=-(\varepsilon_a-\varepsilon_i)f^-_{ia}-\langle\phi_a|\delta v_H|\phi_i\rangle+\langle\phi_a|\delta v^\gamma_{XC}|\phi_i\rangle+\langle \phi_a|y^-_i\rangle+\langle \phi_a|z^+_i\rangle,
\end{equation}
where we defined:  
\begin{equation}
    y^\pm_i(r) = \sum_{jb} f^\pm_{jb}(r) u^\gamma_{ij}(r)
\end{equation}
\begin{equation}
    z^\pm_i(r) = \sum_{jb} f^\pm_{jb}(r) u^\gamma_{ib}(r),
\end{equation}
and the action of $u^\gamma$ on orbital pairs is written shorthand 
$u^\gamma_{ij}(r)\equiv \int u^\gamma(|r-r'|)\phi_i(r')\phi_j(r')dr'$.  
The Hartree potential is $\delta v_H=\int \frac{\delta n(r')}{|r-r'|}dr' $, and the induced density is 
\begin{equation}
\delta n(r) = 2\sum_{jb} (f^+_{jb}+f^-_{jb})\phi_j(r)\phi_b(r).
\end{equation}
The first order change in the short-range exchange-correlation potential is analogous to the real-time approach:  
\begin{equation}
    \delta v^\gamma_{XC} = \frac{1}{\eta}(v^\gamma_{XC}[n_0+\eta\delta n] - v^\gamma_{XC}[n_0]),
\end{equation}
with a strength parameter $\eta\approx 10^{-4}$.~\cite{Baroni1987} 

The  iterative Chebyshev approach to obtain the full, non-TDA, frequency-resolved dipole-dipole correlation function $\sigma(\omega)$ is obtained as~\cite{bradbury_bse_2023}: 
\begin{equation}
    \sigma(\omega)\propto \omega\langle \tilde{\chi}|\delta(\mathcal{L}-\omega)|\chi\rangle,
\end{equation}
and the delta function is calculated as 
\begin{equation}
    \delta(\mathcal{L}-\omega)=\sum_{n=0}^{N_\text{Chebyshev}} c_n(\omega)T_n(\tilde{\mathcal{L}}),
\end{equation}
where $T_{n}(\tilde{\mathcal{L}})$ is the \emph{n}'th Chebyshev polynomial and $\tilde{\mathcal{L}}$ is a scaled Liouvillian with eigenvalues between -1 and 1.  
The optical absorption is obtained from the residues as: 
\begin{equation}
\sigma(\omega)=\sum_{n=0}^{N_\text{Chebyshev}} c_n(\omega)R_n, 
\end{equation}
with $R_n = \langle \tilde{\chi} |T_{n}(\tilde{\mathcal{L}})|\chi \rangle$,
and $\tilde{\chi}^\pm_{ia} = \pm \chi^\pm_{ia}$. For the Chebyshev coefficients, $c_n(\omega)$, we use simple smoothly-decaying weights.~\cite{Weisse2006,bradbury_bse_2023} 

Compared with the TDA,  our inclusion here of the off-diagonal matrix \emph{B} and the negative-frequency component $f^-$ (and therefore the addition of  ``detransitions'' of negative frequency) doubles the spectral range of $\mathcal{L}$, leading to the doubling in the number of required Chebyshev terms, which is typically $1000$ now.

\subsection{Deterministic/Fragmented-Stochastic calculation of long-range exchange}
The dominant cost of the spectra calculation is computing the action of $u^\gamma(|r-r'|)$ on orbital pairs, i.e., the exchange part in TDHF (time-dependent Hartree-Fock). To reduce the cost, we use a mixed deterministic/fragmented-stochastic approach, as developed in Ref. ~\cite{neargap_2023}, which is briefly overviewed below.

As $u^\gamma(r,r')=u^\gamma(|r-r'|)$, we exploit the convolution form of the integrals $u^\gamma_{ij}(r)$ and $u^\gamma_{ib}(r)$: 
\begin{equation}
\label{vwij_pre}
   u^\gamma_{ij}(r)=  \mathcal{F}^{-1} \Big\{ u^\gamma(k) \langle k | \phi_i \phi_j\rangle  \Big\}
\end{equation}
 where $\mathcal{F}^{-1}$ denotes an inverse FFT.  To reduce the effort in Eq. (\ref{vwij_pre}) the interaction $u^\gamma(k)$ is split between long-wavelength, low-\emph{k} components that are included deterministically, while the remainder high-\emph{k} terms are represented through a small sparse-stochastic basis with a constant number of terms (around $1000-2000$) independent of system size. To achieve this fragmentation of $u^\gamma(k)$, the following identity is introduced: 
\begin{equation}
\label{identity}
I = \sum_{k_{\rm{low}}} |{k_{\rm{low}}}\rangle\langle{k_{\rm{low}}}| + \sum_{k_{\rm{high}}} |{k_{\rm{high}}}\rangle \langle{k_{\rm{high}}}|.
\end{equation}
In this basis, the Coulomb interaction is 
\begin{equation}
\label{vW_frac}
\begin{split}
    u^\gamma = &\sum_{\klow}|\klow  \rangle u^\gamma(\klow) \langle\klow| \\
          &+\sum_{\kh} \sqrt{u^\gamma(\kh)}|\kh \rangle \langle \kh| \sqrt{u^\gamma(\kh)}.
\end{split}
\end{equation}
Note that for simplicity we present for the case where all $u^\gamma(k)$ are positive. (In practice some elements are negative due to the use of the Martyna-Tuckerman procedure \cite{MartynaTuckerman1999} to avoid grid-reflection effect, so our actual simulations use the general formulation, as detailed in Ref.~\cite{neargap_2023}.)  

We introduce now a sparse-stochastic basis $\{|\alpha\rangle\}$  (with $N_\alpha$ members) for the $\kh$ space: 
\begin{equation}
    \alpha(\kh) = \pm \sqrt{\frac{N_{\kh}}{S}} A_{\alpha}(\kh)
\end{equation}
where $N_\kh$ is the number of high-$k$ terms, \emph{S} is the length of each fragment, while $A_\alpha$ randomly projects onto a fragment of the $\kh$-space, i.e., it is one within the length-$S$ fragment $\alpha$, and vanishes outside. In principle, for the real-time segment of the paper, we could resample the random projection in the stochastic basis at each time-step to further reduce the stochastic error. However, the high-$k$ contribution is sufficiently small such that the computational cost of resampling the stochastic basis at each time-step is not worth the reduction in error it would give.

The high-$k$ contribution of $u^\gamma$  uses then $N_{\alpha}$ states $|\zeta\rangle$, with components
\begin{equation}
    \langle \kh | \zeta \rangle = \sqrt{u^\gamma(\kh)}  \ \alpha(\kh).
\end{equation}
A full auxiliary basis with $N_\xi = N_{\klow} + N_{\alpha}$   components  is then defined as 
\begin{equation}
    |\xi\rangle = \{\sqrt{u^\gamma(\klow)} \ |\klow\rangle\} \oplus \{|\zeta\rangle\},
\end{equation}
and the interaction is finally a sum of separable terms
\begin{equation}
    u^\gamma=\sum_{\xi}|\xi\rangle \langle\xi|.
\end{equation}

\section{Results}
We first benchmark the resulting TDGKS method on naphthalene and fullerene and then study a chlorophyll a monomer (Chla) and a chlorophyll hexamer reaction-center complex found at the center of Photosystem II (RC-PSII). The Chla model has a methyl acetate ligand
in place of the phytyl chain. The optimized coordinates for both dye systems are taken from Ref.~\cite{Forster_ADF_Chromo_2022}. For Chla, we focus on the visible wavelength region $Q$ absorption bands that are mostly represented by the HOMO to LUMO transition. These excitations are characteristic of the magnesium-center metalloporphyin ring, and depend on the laser polarization direction.~\cite{Gouterman_1961}

The linear-response calculation requires input from a prior ngH-DFT calculation. Table 
\ref{tab:grid} provides the maximum parameters for the ground-state calculations of the four systems. The active space, represented by $N_v$:$N_c$, denotes the number of near Fermi level valence and conduction bands that are explicitly included in both the hybrid DFT calculation and subsequent TD calculation. 

For all systems, the plotted spectra are obtained from the frequency-resolved Chebyshev approach. We verified that the real-time and frequency-based approaches give identical spectra for naphthalene, fullerene, and Chla. The resolution is fixed by setting the spectral width to the number of Chebyshev terms, $\delta\mathcal{L}$/$N_\text{Chebyshev}$, constant, with an energy broadening of roughly 0.2 eV. The real-time calculations were propagated with a time-step of $dt = 0.25$ a.u. for a total number of time-steps $nt = 5000$.

\begin{table*}
\caption{HOMO-LUMO gaps for naphthalene, fullerene, chlorophyll a (Chla), and a 476 atom Photosystem II hexamer dye reaction center (RC-PSII). Also shown are the number of grid-points and occupied states, the maximum number of valence and conduction states, and the range-separation parameter. All energies are in eV.}
\begin{tabular}{|C{2.4cm}|C{0.6cm}C{0.6cm}C{0.6cm}|C{1.0cm}C{1.0cm}C{1.0cm}|C{2.2 cm}|C{2cm}|C{2.2cm}|}
\hline
%\centering
System & $N_x$ & $N_y$ & $N_z$ & $N_{o}$ & $N^{\rm max}_{v}$ & $N^{\rm max}_{c}$ & Optimal $\gamma$ ($\text{Bohr}^{-1}$) & LDA-DFT  & ngH-DFT  \\ 
\hline
Naphthalene & 48 & 44 & 24 & 24 & 24 & 488 & 0.285 & 3.34  & 8.63            \\ 
Fullerene & 60 & 60 & 60 & 120 & 120 & 480 & 0.189 & 1.63  & 5.42            \\
Chla & 84 & 76 & 64 & 116 & 116 & 396 & 0.160  & 1.40 & 4.37          \\ 
RC-PSII & 120 & 148 & 128 & 660 & 200  & 400 & 0.120 & 1.23         &    3.82          \\ \hline
\end{tabular}
\label{tab:grid}
\end{table*}
% put full nvnc hexamer once done

\begin{table}
    \begin{tabular}
    {|C{2.4cm}|C{2.0cm}|C{3.2cm}|}
    \hline
    System & Optical gap & Reference \\
    \hline
    Naphthalene & 4.68 & 4.52 (NWChem)  \\
    Fullerene & 3.33 & 3.30 (NWChem) \\
    Chla & 2.01, 2.20 & 1.99, 2.30 (Expt.) \\
    RC-PSII & 2.20 & 1.95 (Expt.) \\ \hline 
    \end{tabular}
    \caption{Optical gaps (first peak in spectrum) of naphthalene, fullerene, Chla, and RC-PSII using $N_v^{\text{max}}$:$N_c^{\text{max}}$ states. $Q_y$,$Q_x$ peaks are given for Chla. Comparison to available literature values are included for dye systems. All energies in eV. Reference values for naphthalene and fullerene are from LR-TDDFT calculations with NWChem software Ref.~\cite{NWChem2020}, Chla and RC-PSII are experimental values from Refs.~\cite{Chlagasphase_original_2019,Forster_ADF_Chromo_2022}}
    \label{tab:gaps}
\end{table}

\begin{table}
    \begin{tabular}
    {|C{2.4cm}|C{2.0cm}|C{2.0cm}|C{2.0cm}|}
    \hline
    $N_v$:$N_c$ & $Q_y$ & $Q_x$ & $\Delta_{Q_y-Q_x}$ \\
    \hline
    20:40 & 2.23 & 2.45 & 0.22 \\
    40:80 & 2.17 & 2.34 & 0.17 \\
    60:120 & 2.12 & 2.31 & 0.19 \\
    116:198 & 2.09 & 2.26 & 0.17 \\ 
    58:396 & 2.06 & 2.26 & 0.20 \\ 
    116:396 & 2.01 & 2.20 & 0.19 \\ \hline 
    \end{tabular}
    \caption{Convergence of the gap between the $Q_y$ and $Q_x$ bands of Chla with $N_v$:$N_c$. All energies in eV.}
    \label{tab:qxy}
\end{table}

In Table \ref{tab:gaps}, the optical gaps of naphthalene and fullerene are compared with those calculated with the NWChem software package.~\cite{NWChem2020} All optical gaps correspond to the first peak in the absorption spectrum. The NWChem optical gaps are obtained by selecting the lowest root excitation energy with non-zero dipole oscillator strength. The LR-TDDFT NWChem calculations use an aug-cc-pvdz Gaussian basis set and the BNL exchange-correlation functional (with the same range-separation parameter $\gamma$). Good overall agreement is obtained between our new method and the established software for these two hydrocarbon systems.

Figures 1 and 2 show convergence of the optical gaps of naphthalene and Chla with respect to the explicit number of near-gap states included in the spectral calculation. For naphthalene, fullerene, and RC-PSII we present the averaged spectrum, while the Chla peaks are specifically chosen from the x- and y- polarized spectra.   

The convergence of the naphthalene optical gap, Fig. 1, depends more heavily on the inclusion of the virtual orbitals as opposed to the occupied orbitals. However for Chla, Fig. 2, convergence of the $Q_y$ and $Q_x$ peaks requires that all occupied states and a large number of virtual states are included. For these two systems, the optical gap peaks are both converged with the maximum number of conduction states, provided in Table I, and a larger conduction basis did not influence the optical gap peak position. Table \ref{tab:qxy} provides the $Q_y$ and $Q_x$ peak positions as well as absolute differences for Chla with $N_v$:$N_c$ basis size. The distance between the peaks, $\Delta_{Q_y-Q_x}$, is well converged, within 0.03 eV, already for $N_v=20$:$N_c=40$. 

Fig. 3a shows that the optical gap of fullerene  (C$_{60}$) is converged within an error of 0.03 eV for $N_v=60$:$N_c=120$ to $N_v=120$:$N_c=480$. It is known that the optical gap of fullerene is faint,~\cite{c60_ref_specs1998} so to resolve the peak, we use a total of 4000 Chebyshev polynomials. 

While the convergence tests above were for the desired quantity, the low-frequency spectrum, an interesting question is how does the real-time dipole itself (not just the frequency-resolved spectra) converges with basis-set.   Fig. 3b shows this convergence for fullerene, comparing $N_v=80$:$N_c=160$ and $N_v=120$:$N_c=480$. The signal for the smaller subspace accurately captures low frequency oscillations even though the signals do not exactly match.

Moving to the largest system,  Fig. 4 shows the absorption spectrum of the large dye, RC-PSII, as well as the convergence of the optical gap. Unlike the smaller systems studied here, the optical gap of this much larger system converges rapidly with the fraction of valence-conduction states, with $N_v=150$:$N_c=300$ and $N_v=200$:$N_c=400$ giving nearly identical spectra.

With regards to stochastic error, we emphasize that stochastic vectors are only introduced for computing part of the exchange interaction, and all other calculations involved are done deterministically. The use of a sparse-stochastic basis for the high-$k$ component of the exchange interaction, in both the preparation of the ground-state GKS orbitals and the linear-response spectra calculation, introduces only a very small stochastic error. For the four systems, we used $k_{\text{cut}}=1.8, 1.1, 0.9, 0.5$ a.u. respectively, and with the $\gamma$ values provided in Table I, the exchange kernel, $v_\gamma(k) \propto \exp(-k^2/4\gamma^2)/k^2$, is numerically tiny for the high-$k$ stochastically sampled space.

The chosen $k_{\text{cut}}$ values correspond to $N_{k_{\text{low}}}\simeq 5,000$ low-\emph{k} elements. Each system also uses $N_\alpha=1,000$ sparse vectors for the high-\emph{k} space. An overall auxiliary basis of $N_\xi\simeq 6,000$ is then used, irrespective of system size. The stochastic error in our TDDFT calculations is then negligible, much less than 0.01 eV. A more detailed breakdown of the stochastic error is given in Ref.~\cite{neargap_2023}. % need to do quick runs to verify claim

\begin{figure} 
\includegraphics{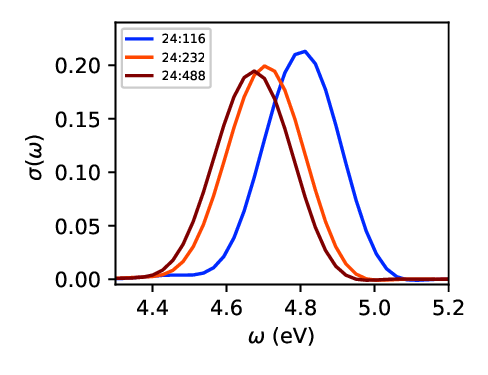}
\label{fig:naph_final}
\caption{Convergence of optical gap of naphthalene with respect to conduction basis $N_c$. The labels correspond to $N_v$:$N_c$. The frequency-domain approach spectra is used here and throughout the other figures, but we confirmed that the real-time spectra are identical.}
\end{figure}

\begin{figure}
\includegraphics{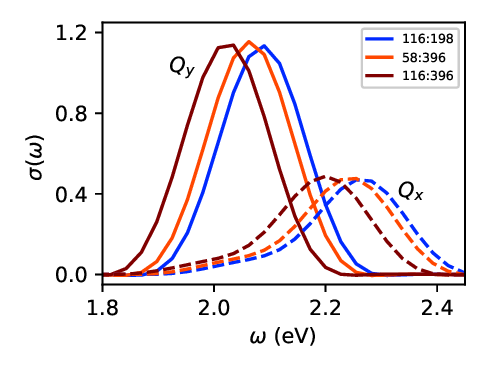}
\label{fig:Chla+xy}
\caption{Convergence of the $Q_y$- and $Q_x$- optical absorption bands of Chla with active space size $N_v$:$N_c$.}
\end{figure}

\begin{figure}
\centering
\begin{subfigure}{.5\textwidth}
  \centering
  \includegraphics[width=1.0\linewidth]{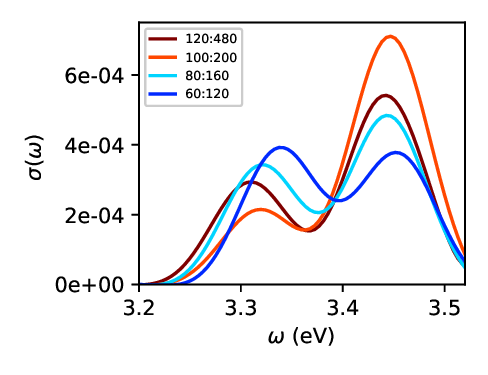}
  \label{fig:sub1}
\end{subfigure}%
\begin{subfigure}{.5\textwidth}
  \centering
  \includegraphics[width=1.0\linewidth]{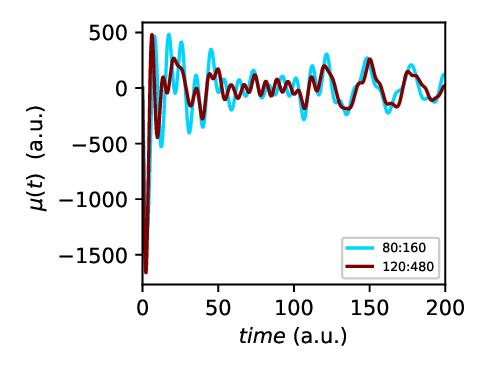}
  \label{fig:sub2}
\end{subfigure}
\caption{(Left) Convergence of optical gap of fullerene with respect to $N_v$:$N_c$. (Right) Time-dependent dipole signal comparison for fullerene using $N_v=60$:$N_c=120$ and $N_v=120$:$N_c=480$.}
\label{fig:test}
\end{figure}

\begin{figure}
  \label{fig:hexamer}
  \centering
  \includegraphics[width=.9\linewidth]{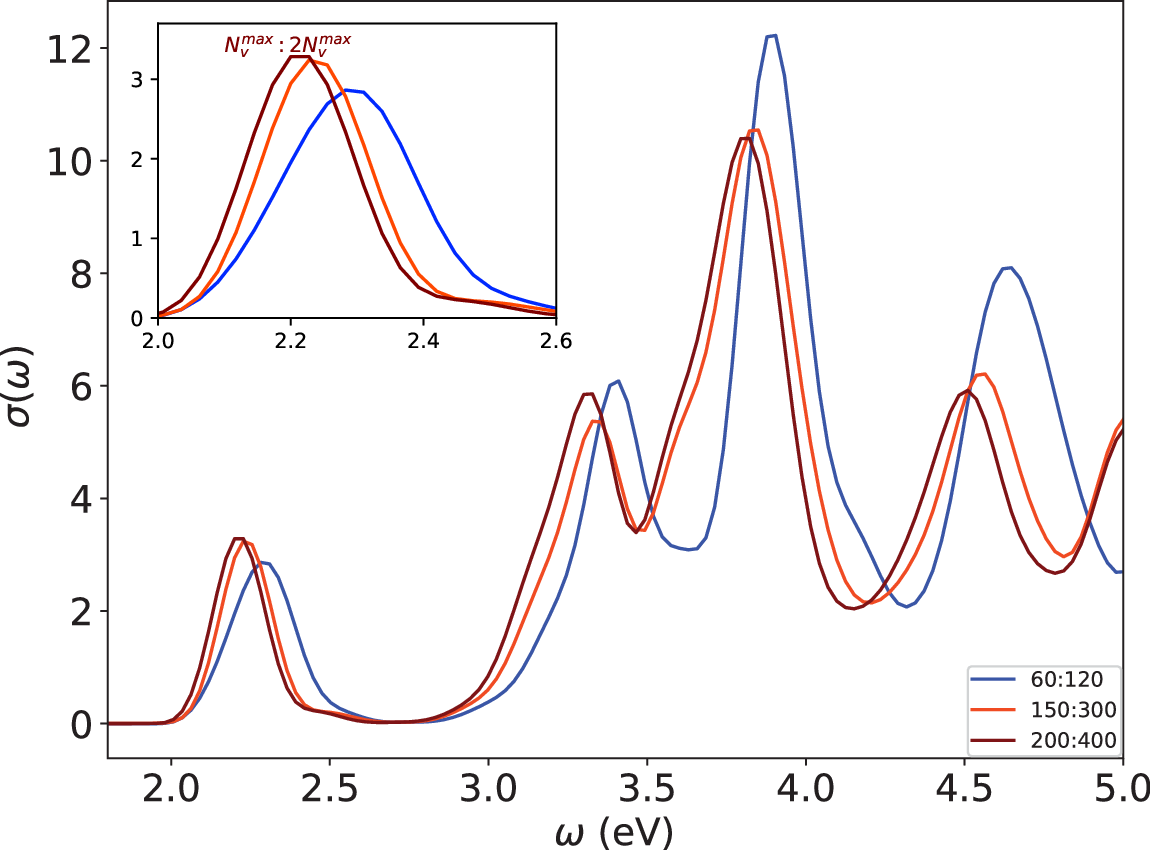}
  \caption{Absorption spectra of RC-PSII with respect to active space size $N_v$:$N_c$. The inset shows the convergence of optical gap with $N_v$:$N_c$.}
\end{figure}

\section{Discussion}
We presented here a GKS-TDDFT formalism for optical spectra that efficiently calculates long-range exact exchange by a mixed deterministic/fragmented stochastic reciprocal-space grid approach. Both a real-time and a frequency domain formulation were studied. These approaches converge the optical gap readily with the number of valence MOs, especially as the system gets larger. Thus, for the largest system, RC-PSII, it is enough to use $N_{v}=150$ rather than the total number of occupied states $N_o=660$. Additionally, the dimension of the conduction basis for the converged hexamer calculation is even smaller than the number of occupied orbitals. The number of required near-gap basis functions decreases relative to the number of occupied electrons as the system size grows due to the response becoming increasingly concentrated near the Fermi energy.

The cost of calculating the exchange matrix elements is reduced in the present approach to $\sim M^2 N_{v} N_{\xi}$, where $M=N_v+N_c$.~\cite{neargap_2023} For the optical gap calculations we use a constant number of $N_{\xi}$, and $N_v$ and $M$ do not increase linearly with number of total MOs in the system. The spectral calculations are even less sensitive to the stochastic basis size $N_{\alpha}$ than are the ground-state GKS calculations. The combination of all of our techniques thus culminates in a method that is particularly well suited to extract the optical properties for systems with a large number of electrons.

To converge the optical gap in the naphthalene and Chla simulations, more conduction states are needed, compared to valence states. When $N_v<N_o$, the core correction to the exchange accounts for much of the effect of the excluded occupied states, by shifting the peak positions with a rigid scissor shift. There is no such equivalent formalism to address the effect of the missing high-lying conduction states. In the future, we will attempt to improve the convergence with respect to the conduction states by working in a pair natural orbital basis to maximize the transition dipole for a cheaper ALDA-TDDFT calculation.~\cite{pno}

The convergence with respect to the selected active space should be significantly improved when extending this approach to the \emph{GW}-BSE method because the screened Coulomb interaction is much weaker than the exchange interaction in the current TDGKS formalism. Specifically, the inclusion of exact exchange in TDGKS yields a better description of charge-transfer excitations than in TDDFT with local or semi-local functionals; however, one needs explicit inclusion of the electron-hole attractive interaction for an even more accurate absorption spectrum. Our recent formulation of the stochastic \textit{GW}-BSE recasts the expensive screened exchange term to a translationally-invariant fitted exchange interaction and a small difference that is sampled stochastically.~\cite{bradbury_bse_2023} This shifts the bulk of the computational effort to calculating the convolutions, as in TDHF, for which we substantially improved the efficiency here. 

In addition to the extension of this LR-TDDFT approach to the BSE, there will be several other developments/extensions of the presented method:  

Fragmented stochastic exchange could be formulated within an orthogonal-projector augmented waves (OPAW) framework. This would extend our recent LDA-OPAW-TDDFT to hybrid functionals with long-range exact exchange.~\cite{nguyen2023timedependent} OPAW enables use of coarser grids and a lower kinetic-energy cutoff compared to norm-conserving pseudopotentials, which will unlock even larger system-size calculations. 

Another pursuit for fragmented-stochastic GKS-TDDFT will be excited-state nuclear gradients. Previous works computing ionic forces in TDDFT were mostly limited to LDA functionals, and the presented method significantly reduces the cost of exchange making long-RSH functionals more accessible.~\cite{Zhang2015SubspaceFO} Note also a recent work with local hybrids.~\cite{furche_2019} 

Going beyond the linear-response regime, the real-time approach would be particularly suited for calculating the hyperpolarizability of molecules as well as simulating strong field phenomena, such as strong field ionization and high harmonic generation. Even though most practical calculations are done on small systems, large grids are required to resolve the electron density when it is far from equilibrium. We expect to see then a different pattern of behavior in converging both the $N_v \oplus N_c$ subspace as well as the parameters in splitting the long-range Coulomb interaction.

\section*{Acknowledgements}
This work is supported by the U.S. Department of Energy, Office of Science, Office of Advanced Scientific Computing Research, Scientific Discovery through Advanced Computing (SciDAC) program under Award Number DE-SC0022198. Computational resources for simulations were provided by both the Expanse cluster at San Diego Supercomputer Center through allocation CHE220086 from the Advanced Cyberinfrastructure Coordination Ecosystem: Services \& Support (ACCESS) program, ~\citep{boerner2023access} and resources of the National Energy Research Scientific Computing Center, a DOE Office of Science User Facility supported by the Office of Science of the U.S. Department of Energy under Contract No. DE-AC02-05CH11231 using NERSC award BES-ERCAP0029462. NCB acknowledges the NSF Graduate Research Fellowship Program under grant DGE-2034835. 

\bibliography{main}% Produces the bibliography via BibTeX.

\end{document}